\documentclass[conference]{IEEEtran}

\usepackage{cite}
\usepackage{amsmath,amssymb,amsfonts}
\usepackage{algorithmic}
\usepackage{graphicx}
\usepackage{textcomp}
\usepackage{xcolor}

\begin{document}

\title{A Trust-Based Approach for Volunteer-Based Distributed Computing in the Context of Biological Simulation}

\author{\IEEEauthorblockN{Sven Hofmann}
  \IEEEauthorblockA{\textit{Department of Computer Science} \\
    \textit{Christian-Albrechts-Universität zu Kiel}\\
    Kiel, Germany \\
    sven.hofmann@stu.uni-kiel.de}
}

\maketitle

\begin{abstract}
  As simulating complex biological processes become more important for modern medicine, new ways to compute this increasingly challenging data are necessary.
  In this paper, one of the most extensive volunteer-based distributed computing systems, called folding@home, is analyzed, and a trust-based approach is developed based upon it.
  Afterward, all advantages and disadvantages are presented.
  This approach uses trusted communities that are a subset of all available clients where they trust each other.
  Using such TCs, the system becomes more organic and responds better to malicious or malfunctioning clients.
\end{abstract}

\begin{IEEEkeywords}
  Grid Computing, distributed, trust, reputation, folding@home, Computational Trust, Trusted-Desktop-Grid, Trust Communities
\end{IEEEkeywords}

\section{Introduction}
At the time of writing, the world is dominated by a worldwide pandemic called COVID-19.
Developing a vaccine against it is one of the most important possibilities to fight this virus.
Nevertheless, time is rare, as the pandemic already caused more than 2.15 million deaths\cite{Worldmeter2021} in the last 12 months.
To speed up this process, vast processing power is needed to simulate the folding of the virus proteins.
This simulated folding process helps scientists in finding new possibilities for a vaccine.
However, this processing power amount cannot be reached with a single supercomputer or a server farm without enormous costs.
To solve this problem, a more sophisticated approach can be applied: volunteer-based distributed computing.
It is called volunteer-based Distributed Computing, which is used by the folding@home project.
By combining the idle power of a large portion of computers worldwide, enormous processing power can be formed.
At the time of writing, the combined processing power reached 0.22 ExaFLOPS\cite{Stats2021}.
In peak time (2020-03), the processing power even exceeded 1.5 ExaFLOPS \cite{Shilov2020} which is even higher than the currently fastest computer globally with about 0.44 ExaFLOPS\cite{Top5002021}.
Furthermore, in contrast to the world's fastest computer, the folding@home project does not need to take all its clients' operating costs into consideration because this cost is donated by the participants who get credits in exchange.

In the following, we introduce the concept and architecture behind the folding@home project.
In contrast, we present a trust-based approach and discuss whether such an approach including trust communities is applicable to a volunteer-based Distributed Computing system like folding@home.

\section{Related Work}
This section gives an overview of Grid Computing, how the folding@home project uses it, and how a Trusted-Desktop-Grid relates to it.

\subsection{Grid Computing}
Grid Computing integrates many computers into a single unit, supported by a robust high-speed network.
Grid Computing can solve complex problems and large amounts of data.
There are five different Grid Computing methods, each of them specialized for a particular type of problem.
(1) \textit{Distributed Supercomputing} is used to solve large and complex problems a single machine would not be able to. Multiple high capacity resources are used for this method.
(2) \textit{High Throughput Computing} uses a large number of CPU cores which can process multiple tasks in parallel and makes it possible to process a large number of small tasks in a short time.
(3) \textit{On-demand Computing} enables resources to be accessible through the grid.
(4) \textit{Data-Intensive Computing} multiple systems share the amount of data to be processed.
(5) \textit{Logistical Networking}, similar to warehouse logistics, schedules - compared to traditional networks - the transport and storage of data inside the grids network.
\cite{Guharoy2017}

The Trusted-Desktop-Grid approach analyzed below in (\ref{tdg}) uses the Distributed Supercomputing method.

\begin{figure}
  \centering
  \includegraphics[width=0.80\linewidth]{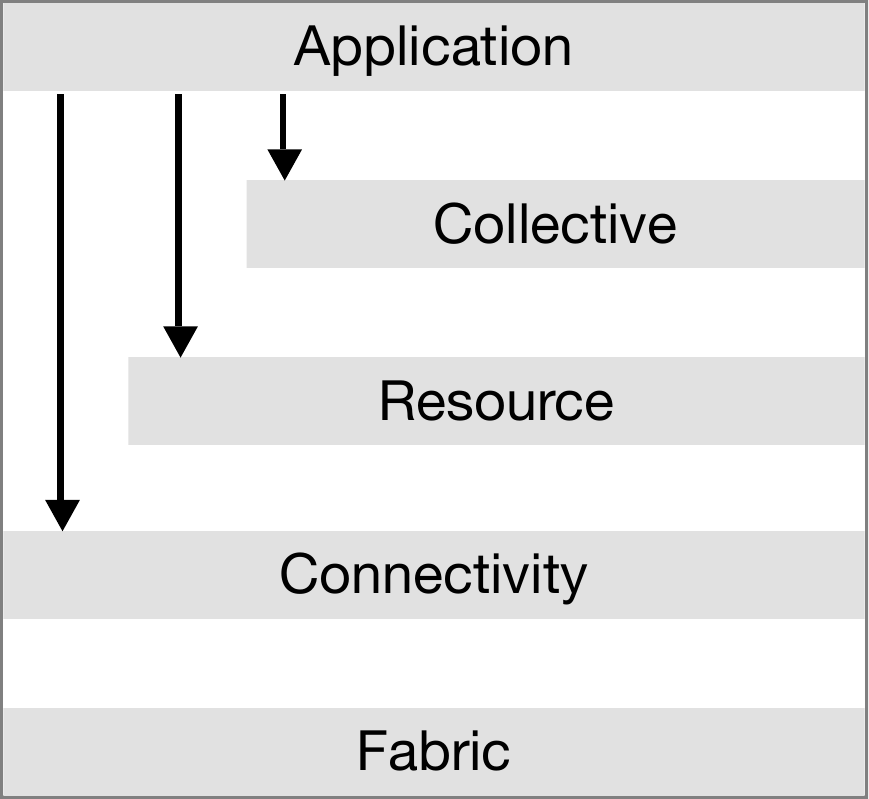}
  \caption{The architecture of Grid Computing (based on \cite{Guharoy2017})}
  \label{arch}
\end{figure}

The architecture of Grid Computing is made out of five layers. (Fig \ref{arch})
First, the \textit{fabric layer} provides the resources to be shared, such as computational or network resources and storage systems.
On top of that is the \textit{connectivity layer} which controls the communication inside the grids network.
It makes use out of the fabric layer.
Next is the \textit{resources layer} which is responsible for monitoring the grid and its resources.
This is followed by the \textit{collective layer} with its APIs and SDKs that provide access to the resources.
It is directly used by finally the \textit{application layer}.
This last layer contains all the applications that make use of the grid. \cite{Guharoy2017}

Scientific research is one area that uses Grid Computing, as scientists often face complex problems that can not be solved with a single machine.
Furthermore, it enables them to work with large amounts of data.

Another solution similar to Grid Computing is Cloud Computing.
The difference to Grid Computing is that computing capabilities come from a computer infrastructure provided by a company via TCP/IP.
Therefore Cloud Computing provides systems dedicated to this task. In contrast, the Grid Computing solution systems share their unused resources \cite{Guharoy2017}.
That dedication has the advantage that there is always the same amount of computing power.
As with Grid Computing, the amount varies over time.
In contrast, Grid Computing uses only the resources that already exist and would otherwise be left unused.
Both solutions have their advantages and disadvantages, and choosing depends on the central operation context.
This article focuses on Grid Computing and adds a trust- and volunteer-based approach to it.

\subsection{The folding@home project}
\label{fah}
The idea of folding@home is to use volunteer-based Grid Computing for simulating the folding of proteins which is a process of self-assembling and is related to many diseases.
It depends on whether a protein folds or misfolds, leading to "disrupting ordinary cellular functions" \cite{Beberg2009}.
The simulation of protein folding is so complex that a single-core CPU can only simulate around 20 nanoseconds of the molecular process per day.
However, the whole simulation takes from milliseconds to seconds, making it impossible to compute for a single computer or cluster \cite{Beberg2009}.
Moreover, the costs for such a cluster would be enormous.

Grid Computing is an excellent solution to this problem.
There are hundreds and thousands of computers around the globe, many of them just running idle.
These machines are controlled by many independent people and not by one organization.
They have to be convinced to contribute voluntarily to the project.
One solution to this is gamification.
Every contributor gets credits for the work units his machine has completed.
The folding@homes statistic website lists all the users ranked by the credit points earned so far.
The participants can additionally form teams to accumulate their credit points.
More prominent organizations can use this reputation system for better social standing.
Gamification also comes with some disadvantages, like people running the client software on someone else's computer to gain more credit points.
According to \cite{Beberg2009} both "installing the clients on machines they do not own at school or work" and "the use of Trojan horses on P2P file sharing systems to install the client and gain in the statistics" have been taken into consideration, and these problems are treated by banning the participants as well as deleting their scores.

All of this leads to volunteer-based Grid Computing.
In the case of folding@home, this grid has a central control instance.
Every volunteer downloads the \textit{client software} from the webserver and installs it on his local machine.
The client provides some settings like how much of the computer's performance should be used and whether to do it when the computer is idle.
Then the client begins to ask the \textit{assignment server} for work that has to be done.
The result is the address of a \textit{work server} which provides the binaries, also called work unit, to compute.
Each work unit has to be compatible with the computer's architecture and operating system.
Providing binaries as work units makes it very flexible for the task to be done, as this binary can contain any algorithm as long as it is compatible.
The client itself has no information about the tasks hardcoded into it.
The work unit's result is then sent back to the work server, or, if this server is not available, to a \textit{collection server} which acts as a buffer and sends the information to the work server as soon as it comes back online.
The work server computes the results, shares them with the \textit{web and stats server}, and determines the next work unit to be done.
This flow works as long as the client machine is running.
However, as the client software runs in the background, the participant can shutdown his machine at any time, or it might even crash.
Therefore, the results of the work unit have to be stored recurrently \cite{Beberg2009}.

At the time of writing, the most important diseases related to protein folding is COVID-19.
This world-dominating pandemic let "the project grew from $\thicksim$30,000 active devices to over a million devices around the globe" \cite{Zimmerman2020}.

\begin{figure*}[ht!]
  \centering
  \includegraphics[width=0.6\linewidth]{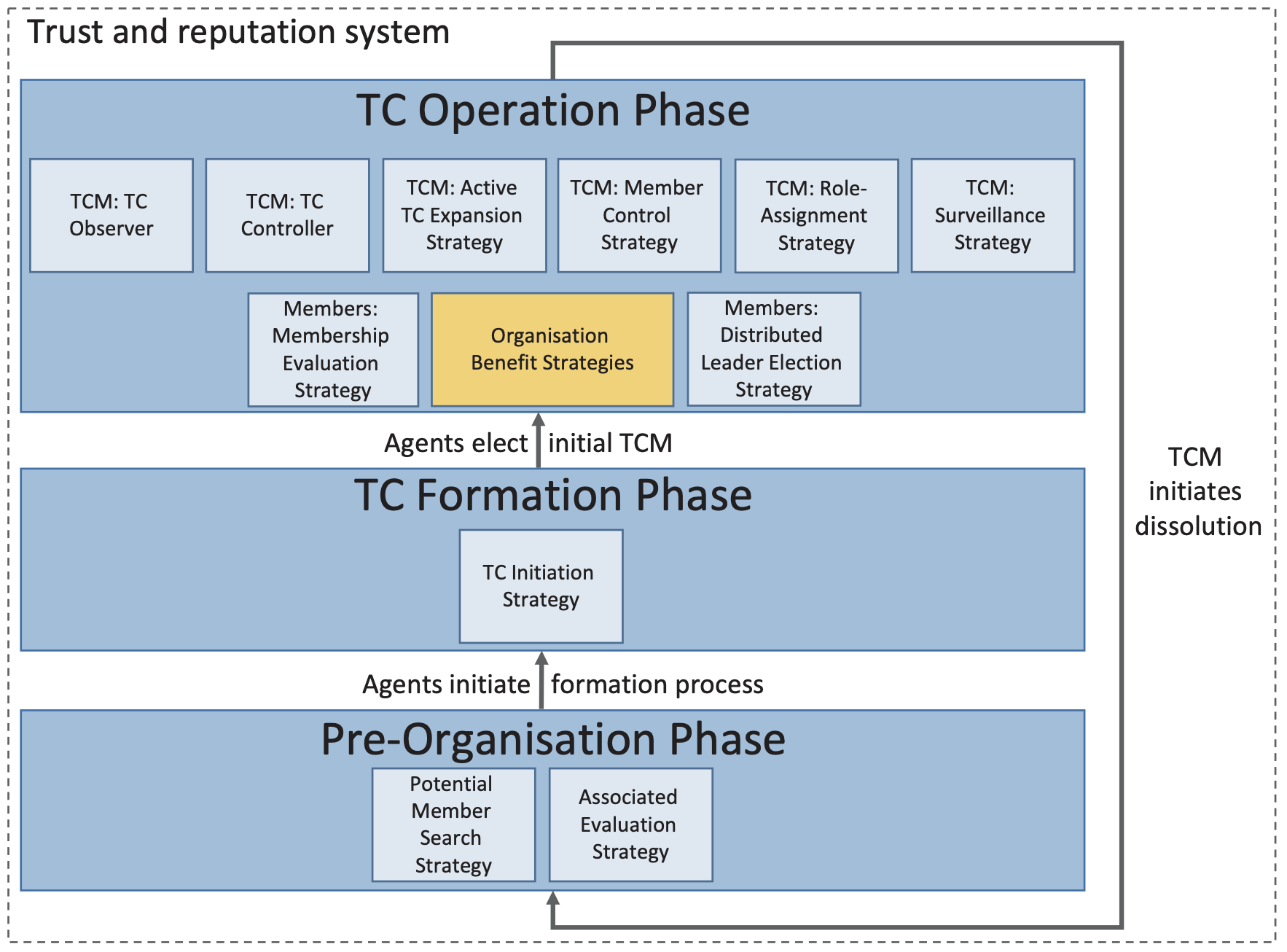}
  \caption{"The life-cycle of an eTC: During the pre-organisation phase, potential members are searched. Then, the eTC is formed (TC formation phase). Afterwards, a TCM is elected and the eTC is in the TC operation phase, where the TCM and the members use strategies, e.g. to observe the environment and control the eTC." \cite{Edenhofer2016}}
  \label{trust-reputation}
\end{figure*}

\subsection{Trusted-Desktop-Grid}
\label{tdg}
A Trusted-Desktop-Grid (TDG) at its core is similar to Grid Computing - with the addition of a trust-based selection method.
Like Grid Computing, it consists of a set of agents that collectively work towards a common goal.
However, in contrast, a TDG does not need a central server that determines which agent should do which work unit.
A TDG acts more organic by selecting an interaction partner by their reputation.
Furthermore, it can handle multiple submitters of work units.
Every agent can either accept a work unit by giving a pessimistic deadline or reject it.
Even accepted work units can be rejected afterward.
Based on the work unit result, the participant gets rated.
These ratings $r$ are between $-1$ and $1$, meaning bad and good.
Based on a set of ratings $ R = \{r_1,r_2,r_3\} $ a reputation is calculated using an aggregation function $\tau(R) \in [0,1]$ with $0.5$ being a neutral reputation \cite{Kantert2016}.

\begin{figure}
  \centering
  \includegraphics[width=\linewidth]{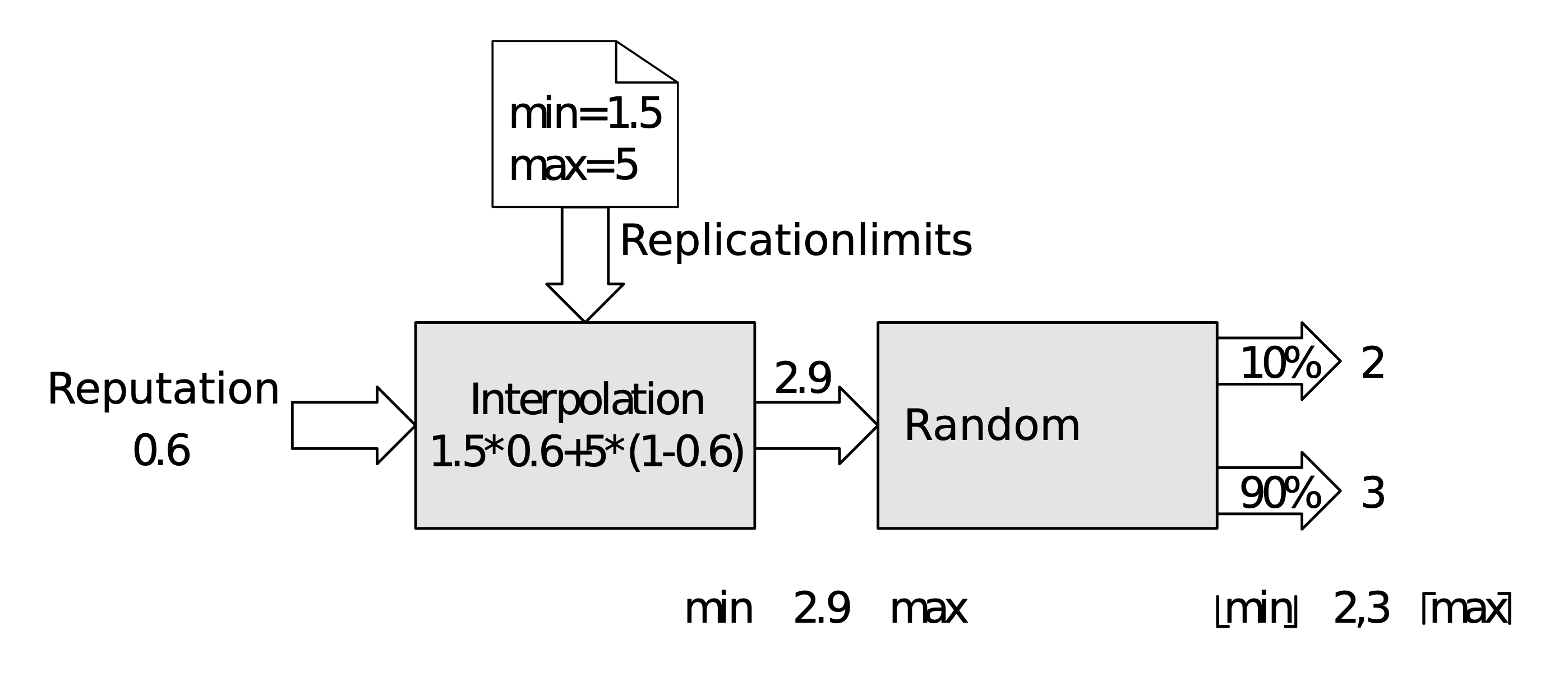}
  \caption{"We calculate the minimum replication factor $f_{min}$ for each worker. First, we interpolate $f_{min}$ based on the reputation $\tau$ between defined limits (here from the interval $[1.5,5]$). Then, we round $f_{min}$ to the next integer using a roulette wheel random generator." \cite{Kantert2016}}
  \label{replication-factor}
\end{figure}

With these reputation values, a minimum replication factor $f_{min}$ can be calculated. (Fig. \ref{replication-factor})
The replication factor describes how many other agents should receive the same work unit to get a trustable result.
An agent with a good reputation will have a low replication factor as this agent's result is trustable and has to be checked by only a small amount of other agents or eventually no agents at all.
For the calculation, it is necessary to define limits for the replication factor.
A higher maximum limit for the replication factor means more reliability but less throughput, as more agents have to work on the same work unit.
The minimum replication factor represents the number of other agents that calculate the same work unit even if the agent has the highest reputation.
If an agent with a miserable replication factor ($\tau(R) < 0.4$) is selected, the replication will be higher.

The minimum replication factor can be used for three distribution strategies.

1) The \textit{Dynamic Random Distribution Strategy (DRDS)} is the most simple of them.
It randomly selects one agent and uses its $f_{min}$ to select the same amount of other agents.
\cite{Kantert2016}

2) The \textit{Dynamic Ordered Distribution Strategy (DODS)} orders the agents by their $f_{min}$ in the first step.
The first work unit is then distributed to the agent with the lowest $f_{min}$ and to all the following until the highest $f_{min}$ of them is satisfied.
The second work unit is then distributed to the next agent with the lowest $f_{min}$ with no work unit \cite{Kantert2016}.

3) The \textit{Dynamic Grouping Distribution Strategy (DGDS)} groups the agents into "trusted" ($\tau > 0.7$), "untrusted" ($\tau \leq 0.4$) or "undecided" ($0.4 < \tau \leq 0.7$).
One "untrusted" agent and up to $\lfloor \frac{f_{min} - 1}{2} \rfloor$ other "untrusted" agents are selected in the initial step.
The same amount of "trusted" agents are selected then.
Lastly, the group gets filled up by "undecided" agents to match the group's highest $f_{min}$.
It ensures that the "untrusted" agents cannot form the majority \cite{Kantert2016}.

\begin{figure*}[ht!]
  \centering
  \includegraphics[width=0.7\linewidth]{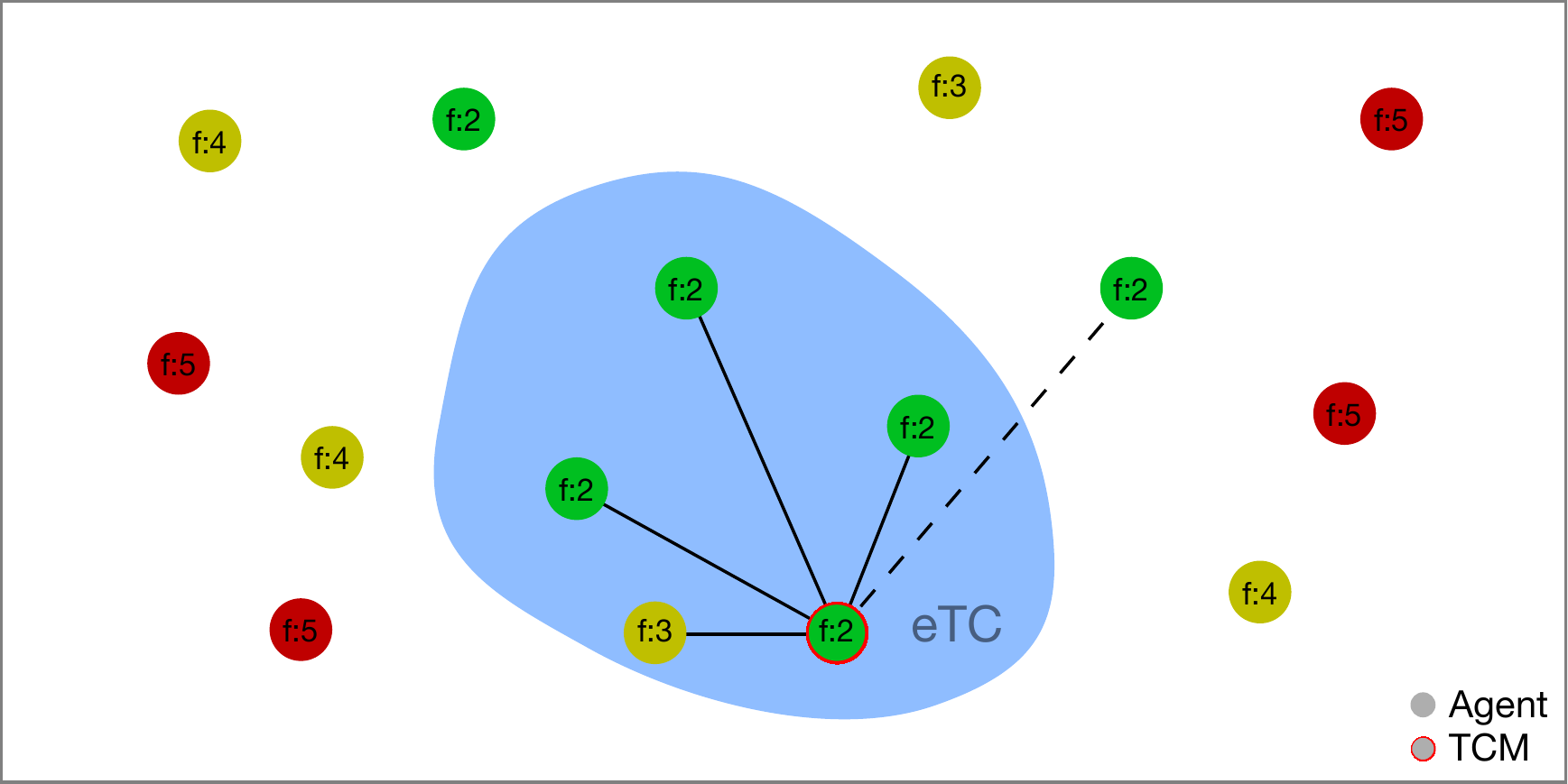}
  \caption{This is an eTC with a TCM distributing WUs to the members and inviting a new agent to join, trying to separate the agents with a good reputation from those with a bad one.}
  \label{etc}
\end{figure*}

\subsection{Trusted-Desktop-Grid with Trust Communities}
A further approach to TDG is the concept of Trust Communities (TC).
A TC is a subset of agents participating in a TDG, where each agent can trust another.
Every member of a TC is free to leave at any time, and new members are selected by their reputation.
In this article, we use explicit TCs (eTC) to compare them to the folding@home project.
Explicit TCs have the specialty to have a Trust Community Manager (TCM) which organizes the eTC.

The eTC has a life-cycle that repeats (Fig. \ref{trust-reputation}).
In the pre-organization phase, the first step of forming an eTC, all the agents are unrated and begin to rate each other based on the work results.
They receive a rating between -1 (e.g., the working unit was rejected) and 1 (e.g., correct and in time).
As soon as a certain number of agents have strong trust relations, they can decide to form an eTC, based on whether they would profit from this eTC or not.
In this TC formation phase, the agents negotiate with all the other potential members because they might not know every single one yet.
During this process and at every other time, all the agents are free to leave the eTC.
If enough agents have decided to stay, they elect the Trust Community Manager (TCM).
This election can be done by criteria like trust, reputation, or availability.
After this election, the operational phase begins.
In this phase, the TCM's organizational tasks are to monitor the community's performance, remove members that are performing worse than before, and let agents join to increase the eTC's performance.
The monitor task is also distributed to the other agents, as it is not possible to monitor all agents by the TCM.
If the benefit of operating inside an eTC is too low, the agents decide to leave the eTC.
As soon as a certain threshold is reached, the eTC gets dissolved by the TCM.
Then, this whole life-cycle begins again \cite{Edenhofer2016}.

\section{Approaches}
This section will summarize the centralized approach of the folding@home project like it is currently used.
After that, we will elaborate on a second approach that uses the previously introduced Trusted-Desktop-Grid combined with an explicit Trust Community.

\subsection{Centralized Distributed Computing}
As mentioned earlier, the folding@home project uses a centralized approach for selecting workers for the work units.
This approach needs a Server-Client architecture with a server that coordinates all the work units and clients that compute the work units.
As mentioned in \ref{fah}, the assignment server balances the clients' incoming work requests between the work servers, which then forward the work units to the clients.
Theoretically, the client has an unlimited amount of time to process the work unit (WU).
When the user shuts his machine down, and the client cannot continue to process, this unlimited amount of time is used.
However, in the usual case, the WU gets processed in a suitable amount of time.
Then the results are returned to the work server.
Based on the WU's complexity, the client receives credits for his work, which are collected on the statistics server.

\subsection{Trust-Based Organic Distributed Computing}
A trust-based based approach for organic Distributed Computing is way more organic.
No strict server-client architecture is needed, but instead, the participants, also called agents, organize themselves using a reputation system.

As already mentioned, we want to take a closer look at the explicit Trust Communities (eTC) and how they can be applied to the folding@home project (Fig. \ref{etc}).
First, we remove the server-client architecture, and every participant of the project becomes an agent.
To make both approaches comparable, the software of the agents, previously known as the clients, has to be limited to only accept work units and not also send new ones to the grid, except they become the Trust Community Manager.
Each organization that was previously hosting a work-server becomes a fully functional agent with no limitations.
We will call agents with the ability to distribute WUs work agents.
Therefore, these agents could also accept work units from other organizations.
To match the credit system of the foldign@home project, we will also give credits for solving WUs.
However, they are shared evenly between all the agents that worked on the same WU.
This sharing of credits makes it preferable to join an eTC as the replication factor is likely lower.
The on average lower replication factor leads to fewer agents working on the same WU, resulting in more credits for solving it.
The higher amount of credits makes it profitable for the agents to join an eTC because their only goal is to gain as many credits as possible.
Aiming for the highest performance, we have to consider that too egoistic agents might try to solve the WUs independently, so they do not need to share credits with other agents.
This behavior has to be prevented, as it lowers the throughput and making the whole system unprofitable.
As a result, we expect the eTCs to consist of the highest-rated agents, increasing the average performance, and even more critical, the throughput.
This results from fewer agents being involved in the processing of one work unit, which increases the parallelism.
To reduce the need for a central statistics server, we are using a blockchain to keep track of the credits.
Every agent can check their balance of credits by himself.

With the new setup, we now initiate the process of forming eTCs.
Each agent randomly distributes its work units to the other agents and rates them afterward by criteria like correctness, time to compute, or rejection.
As a result, the work agents begin to invite the agents with the best reputation to join their eTC.
If enough agents decide to join the eTC, the Trust Community Manager will be elected.
The election can be based on several criteria, e.g., availability or responsiveness.
For simplicity, we will use availability as the main criteria.
Therefore, the work agent that invited the other agents to join the eTC becomes the first TCM.
In case this agent becomes unavailable, the next available agent with the longest time participating in the eTC becomes the next TCM.
The TCM has the role of organizing the eTC.
First, to monitor the performance of all the members and assign other agents the role to monitor.
Second, to look for new members with a good reputation that would increase the eTC's performance.
Third, to remove members who have a decrease in performance.
Lastly, to dissolve the eTC as soon as the project is completed.
In our approach, we expand this role also to distribute the work units.
This new role makes it possible to continue distributing WUs if the former agent becomes unavailable and a new TCM is elected.
To enable this role, we need every member of an eTC to store a copy of the binary needed to construct new WUs.

\section{Comparison}
The newly developed trust-based approach might look slightly similar to the centralized one, but it performs differently.

Imagine an agent/client that receives a work unit and does not process it in order to harm the grid or because its owner shut down the agent/client.
With the traditional approach, this work unit would have to be redistributed to a new client after some threshold.
This redistribution costs additional time, while the new approach can use the other agents' results and penalize the non-responding agent by awarding a negative rating.
With this negative rating, it is less likely that this happens again, in contrast to the centralized approach, where the client can still receive new WUs when it is back online.
When already operating in an eTC, the agent's behavior can result in removal from the eTC, which has a more critical impact on preventing this situation.

Another possible situation is when the work server becomes unavailable, which would cause the traditional approach to come to a complete hold as soon as all WUs are finished.
This approach still can collect the results with the collection servers, but the clients cannot receive new WUs.
However, with the new approach, it is possible to keep the work running.
In response to the work unit distributing TCM becoming unavailable, a new TCM can be elected.
This new TCM continues the distribution of work units and keeps the system running.
The results are then returned to the initial TCM when it is available again.
With this new ability comes a considerable overhead, as more information has to be exchanged between the agents.
This overhead is one of the disadvantages of the trust-based approach.

Selecting the agents with the best reputation to form an eTC makes the higher throughput another advantage of the new approach over the traditional one.
If the WUs are smaller and well parallelizable, the higher throughput becomes even more significant regarding the grid's performance.

A considerable disadvantage of the trust-based approach is the necessary complexity of the agent's software.
The former clients now can have the same task of distributing WUs previously exclusive to the work server.

\section{Conclusion}
We presented the concepts of a Trusted-Desktop-Grid and an explicit Trust Community and used these to develop a trust-based extension for the folding@home project.
This approach shows that trust-based distributed computing is indeed applicable to the folding@home project.
Furthermore, it gives some performance and reliability advantages over the centralized approach, although some overhead will occur as a drawback.
Nevertheless, the new approach could take the abilities of the folding@home project even further.

\bibliographystyle{IEEEtran}
\bibliography{references}

\end{document}